\documentclass[doublecol]{epl2} 
\usepackage{amsmath,bbm,amssymb}
\title{Crossover from weak to strong coupling regime in dispersive circuit QED}
\author{I. Serban\inst{1,2} \and E. Solano\inst{1,3,4} \and F.K. Wilhelm \inst{2}}
\shortauthor{I. Serban \etal}
\institute{
\inst{1}Physics Department, ASC and CeNS,
Ludwig-Maximilians-Universit\"at, Theresienstr. 37, 80333 Munich,
Germany\\
\inst{2}IQC and Dept.~of Physics and Astronomy, University of
Waterloo, 200 University Ave. W, Waterloo, ON, N2L 3G1, Canada\\
\inst{3}Max-Planck Institute for Quantum Optics,
Hans-Kopfermann-Strasse 1, D-85748 Garching, Germany\\
\inst{4}Secci\'{o}n F\'{\i}sica, Departamento de Ciencias,
Pontificia Universidad Cat\'{o}lica del Per\'{u}, Apartado 1761,
Lima, Peru}

\abstract{
We study the decoherence of a superconducting qubit due to the dispersive coupling to a damped harmonic oscillator. We go beyond the weak qubit-oscillator coupling, which we associate with a {\it phase Purcell effect}, and enter into  a strong coupling regime, with qualitatively different behavior of the dephasing rate. We identify  and give a physicaly intuitive discussion of both decoherence mechanisms.
Our results can be applied, with small adaptations, to a large variety of other physical systems, e.~g.~trapped ions and cavity QED, boosting theoretical and experimental decoherence studies.}

\begin{document}
\maketitle

\section{Introduction}
With a thrust from applications in quantum computing, the
manipulation of quantum states in superconducting nanocircuits has
made tremendous progress over the last decade
\cite{Bertet05b,Wallraff04,Vion02,Duty04,Makhlin01,Pashkin03,McDermott05,Oliver05,Plourde05}. A
crucial step for these successes is the understanding of decoherence
and the design of good measurement schemes. The latter is a particular
challenge as the detector is made using the same technology as the
system being  detected i.e.~the qubit. Also, the measurement timescale cannot be considered to be
infinitesimally short as compared to the intrinsic scales of the qubit
evolution. Thus, understanding the measurement process is crucial both
fundamentally and for improving experiments.

A specifically attractive development is the emergence of  circuit
quantum electrodynamics (cQED)
\cite{Buisson01,You03,Goorden04,PRBR033,Kleff04,Yang04,Sarovar05,Mariantoni05},
where effective Hamiltonians, similar to those of the coherent
light-matter interaction of quantum optics and in  particular of
cavity QED, can be realized in the microwave frequency domain. There
are many approaches to realize the qubit, including flux and charge,
and the cavity, including a superconducting quantum interference device (SQUID) or a coplanar waveguide.

In this context, measurement protocols making use of dispersive
qubit-oscillator interactions~\cite{Bertet05b,Wallraff04} are useful for reducing the backaction
on the qubit \cite{Lupascu04}. For example, in the flux qubit--SQUID
combination, as in the Delft setup of Refs.~\cite{Bertet05b, Bertet05c}, the
SQUID behaves like a harmonic oscillator. Its inductive coupling to
the flux qubit leads to a frequency shift depending on the qubit
state $\Omega_{\uparrow,\downarrow}=\sqrt{\Omega^2\pm\Delta^2}$.
Here, $\Omega$ is the bare oscillator frequency and $\Delta$ is the
quadratic frequency shift. A measurement of the SQUID
resonance frequency provides information of the qubit state. While the manipulation of the qubit is usually performed at
the optimum working point \cite{Vion02}, the readout can and should
be performed in quantum nondemolition measurement i.e.~in the pure
dephasing limit.

In this letter we study the decoherence of a qubit
due to the dispersive coupling to a damped harmonic oscillator, taking the Delft setup as an example though our results may be adapted to several  physical systems. In the Purcell effect a narrow oscillator linewidth enhances the absorbtion of the resonant photon emitted by the two-level atom and thus the energy relaxation of the latter. In the weak
qubit-oscillator coupling regime (WQOC), we explain the behavior of dephasing in terms of a similar process, the phase Purcell effect.
This regime is characterized, as we will be show later, by
$\Delta/\Omega<\sqrt{\kappa/\Omega}/
(1+n(\Omega))^{1/4}$, where $n(\Omega)$ is the Bose function  at the frequency $\Omega$ and environment temperature $T$. { The main result of this work lays} beyond the WQOC, in a regime where fast qubit-oscillator entanglement plays the dominant role. We find a qualitatively different behavior of the dephasing rate. 
The divergence of the qubit dephasing rate $1/\tau_\phi\propto 1/\kappa$  when the oscillator decay rate $\kappa \rightarrow 0$ is lifted by the onset of the strong coupling regime.

The Hamiltonian describing the Delft setup~\cite{Bertet05c}
can be written as 
\begin{eqnarray}
\!\!\!\!\!\hat{H}\!\!&=&\!\!\underbrace{\frac{E}{2}\hat{\sigma}_z+\hbar\Omega
\left(\hat{a}^\dagger \hat{a}+\frac{1}{2}\right)+\frac{\hbar
\Delta^2}{4\Omega}(\hat{a}+\hat{a}^{\dagger})^2\hat{\sigma}_z}_{\hat{H}_S}+\hat{H}_{D}
\label{eq:Hamiltonian} . \nonumber \\
\end{eqnarray}
Here, $\hat{a}$ and $\hat{a}^\dagger$ are the annihilation and
creation operators of the harmonic oscillator, $\hat{\sigma}_z$ acts
in the Hilbert space of the qubit and $\hat{H}_D$ describes the
damping of the oscillator. A full-length derivation of Hamiltonian (\ref{eq:Hamiltonian}) and discussion of the approximations used is given in Ref. \cite{Serban06}. It basically derives the Hamiltonian from the equations of motion of the Josephson phases across the junctions and truncates the SQUID potential to the second order.

We will show that key experiments~\cite{Bertet05b, Wallraff04} are
performed outside the WQOC. Moreover, a very recent
experiment~\cite{Schuster06} explicitly relies on the use of a
strong dispersive coupling regime. We
demonstrate that the dephasing rate $1/\tau_\phi\propto 1/\kappa$
for WQOC, and $1/\tau_\phi\propto\kappa$ at
strong coupling. We discuss the crossover between these regimes and
its dependence on $\kappa$ and temperature $T$. We provide physical
interpretations of both regimes, the former as a phase Purcell
effect and the latter as the onset of qubit-oscillator
entanglement. The results of the present study may be extended
straightforwardly to any system with similar dispersive
qubit-oscillator coupling: the charge-qubit--coplanar wave guide system (see Yale
setup~\cite{Wallraff04}), trapped ions~\cite{Leibfried03} and 3D
microwave cavity QED~\cite{Raimond01}, quantum
dots~\cite{Balodato05}, among others.

\section{Method}
In studying the qubit dephasing we are facing the challenge of a complex non-markovian environment consisting in the main oscillator (i.e. SQUID) and the ohmic bath. Moreover, the qubit couples to a non-Gaussian variable of its environment. Therefore the tools developed for Gaussian baths \cite{Weiss99} cannot be applied in this system for arbitrary strong coupling between the qubit and the oscillator. 

We study the qubit dynamics under the Hamiltonian (\ref{eq:Hamiltonian}) for arbitrary $\Delta/\Omega$, assuming essentially
the dimensionless oscillator decay rate $\kappa/\Omega$ as the {\em
only} small parameter. In this regime we avoid  over-damping of the oscillator and the strong backaction on the system which this would cause. We give in the following a brief description of the crucial steps
and approximations of the calculation. We model the damping, associated with the
oscillator decay rate $\kappa$, in the Caldeira-Leggett way by a
bath of harmonic oscillators
\begin{eqnarray}
\!\!\!\hat{H}_D\!\!\!&=&\!\!\!\!\underbrace{\sum_j\hbar\omega_j
\left(\hat{b}_j^{\dagger}\hat{b}_j+\frac{1}{2}\right)}_{\hat{H}_B}
\!+\!\underbrace{\sum_j\frac{\hbar(\hat{a}+\hat{a}^{\dagger})}
{2\sqrt{m\:\Omega}}\frac{\lambda_j(\hat{b}_j^{\dagger}+\hat{b}_j)}
{\sqrt{m_j\:\omega_j}}}_{\hat{H}_{I}}+\hat{H}_c, \nonumber \\\label{eq:bath}
\end{eqnarray}
with $J(\omega)=\sum_j\lambda_j^2\hbar/(2m_j\omega_j)
\delta(\omega-\omega_j)=m\hbar\kappa\omega\Theta(\omega-\omega_c)/
\pi$ and $\hat{H}_c$ the counter term \cite{Ingold98,Cohen92,Nato06II} where $\Theta$ is the Heaviside step
function and $\omega_c$ an intrinsic high frequency cut-off. 
Our starting point is the Born-Markov master equation in the weak coupling
to the bath limit for the reduced density matrix $\hat{\rho}_S$ in the qubit-oscillator Hilbert space
\begin{eqnarray*}
\dot{\hat{\rho}}_S(t)&=& 
\frac{1}{i \hbar}\left[\hat{H}_S,\hat{\rho}_S(t)\right]\label{mastereq}\\
 &+&\int_0^t\!\!\! \frac{dt'}{(i \hbar)^2}\:{\rm Tr}_B \!
\left[\hat{H}_{I},[\hat{H}_{I}(t,t'),\hat{\rho}_S(t)\!\otimes\!\hat{\rho}_{B0}]\right]\!.\nonumber
\end{eqnarray*}
This approach is valid at finite temperatures $k_BT \gg  \hbar \kappa$, for times $t\gg1/\omega_c$ \cite{Alicki06,Nato06II}, which is
the limit we will discuss henceforth. We start from a standard factorized initial state for all 
subsystems. We express $\hat{\rho}_S(t)$
in the qubit basis and represent its elements, which are still oscillator operators, in phase-space as
\begin{equation}
\hat{\rho}_S = \left( \begin{matrix}
\hat{\rho}_{\uparrow\uparrow}&\hat{\rho}_{\uparrow\downarrow} \\
\hat{\rho}_{\downarrow\uparrow}&\hat{\rho}_{\downarrow\downarrow}\end{matrix}\right)
,\:\: \hat{\rho}_{\sigma\sigma'}=\int\!
\frac{d^2\alpha}{\pi}\:\:\chi_{\sigma\sigma'}(\alpha,\alpha^*,t)\hat{D}(-\alpha),\nonumber
\end{equation}
where $\chi_{\sigma\sigma'}$ is the characteristic Wigner function
and $\hat{D}=\exp(\alpha\hat{a}^{\dagger}-\alpha^*\hat{a})$ the displacement operator \cite{Cahill69}. Independent of our work,
Ref.~\cite{Gambetta06} has used a different phase-space
representation to calculate the qubit dephasing rate. We characterize the qubit coherence by
$C(t)=\left\langle\hat{\sigma}_x\otimes
\hat{\mathbbm{1}}\right\rangle=2{\rm
Re\:\:Tr}\hat{\rho}_{\uparrow\downarrow}(t)$ which can be easily
shown to be $C(t)=8\pi{\rm Re}\chi_{\uparrow\downarrow}(0,0,t)$.
After a rather long but essentially straightforward calculation,
one obtains for $\chi_{\uparrow\downarrow}$ a \emph{generalized}
Fokker-Planck equation
\begin{eqnarray}
\!\dot{\chi}_{\uparrow\downarrow}(\alpha,\alpha^*,t)&=&
\Big(\!\!\left(\alpha
(k_1+i \Omega)+\alpha^*k_1\right)\partial_{\alpha}\nonumber \\
&+&
\left(\alpha^*(k_2-i \Omega)+\alpha
k_2\right)\partial_{\alpha^*}\nonumber\\
-\frac{i \Delta^2}{2\Omega}(\partial_\alpha-\partial_{\alpha^*})^2 &+&p(\alpha+
\alpha^*)^2\!\Big)\chi_{\uparrow\downarrow}(\alpha,\alpha^*,t)\label{offdiag},
\end{eqnarray}
where
\begin{eqnarray}
\!\!\!\!\!\!\!  k_{1,2} & = & -\frac{\kappa}{4}\left(2\mp\frac{\Omega_{\uparrow}}{\Omega}(1+2n_{\uparrow})\pm\frac{\Omega_{\downarrow}}{\Omega}(1+2n_{\downarrow})\right) , \\
 p &= & -\frac{\kappa}{8\Omega}\left(\Omega_{\uparrow}(1+2n_{\uparrow})+\Omega_{\downarrow}(1+2n_{\downarrow})\right)-\frac{i \Delta^2}{8\Omega}
\end{eqnarray}
and $n_{\sigma}=n(\Omega_{\sigma})$ is the Bose function. To solve
Eq.~(\ref{offdiag}) we make a Gaussian ansatz for
$\chi_{\uparrow\downarrow}$. 
\begin{eqnarray}
\chi_{\uparrow\downarrow}&=&A(t)\exp(-M(t)\alpha^2-N(t)\alpha^{*2}-Q(t)\alpha\alpha^*).
\end{eqnarray}
This ansatz includes coherent and thermal states. In the following we assume the  oscillator to be initially in a thermal state, in equilibrium with its environment. { This implies $Q(0)=1/2+n(\Omega)$ and $M(0)=N(0)=0$}.  Due to the quadratic (pure dephasing) form of
the Hamiltonian (\ref{eq:Hamiltonian}), obtain a closed system of
ordinary differential equations for the parameters of the Gaussian ansatz, { see also Ref. \cite{Serban06}}. This system can be easily solved perturbatively in $\Delta$ { in the weak coupling regime}, or numerically, { (for arbitrarily strong coupling)}, and we can extract the
dephasing time  $\tau_\phi$ from the strictly exponential
long-time tail of $C(t)=8\pi{\rm Re}A(t)$.

\section{Weak qubit-oscillator coupling} 
{Before solving Eq. (\ref{offdiag}) in a general manner, we revisit the case of small $\Delta$}. Up to the lowest non-vanishing order $\Delta^4$, the
analytically calculated WQOC dephasing rate is
\begin{eqnarray}
\frac{1}{\tau_\phi}=\Delta^4
\frac{n\!\left(\Omega\right)\left(n\!\left(\Omega\right)+1\right)}{\Omega^2}
\left(\frac{\kappa}{\kappa_m^2}+\frac{1}{\kappa}\right),\label{dephasingrate}
\end{eqnarray}
where $\kappa_m= \sqrt{2k_BT\Omega/(\hbar(1+2\:n(\Omega)))}$.
The term $1/\kappa$ exactly reproduces the Golden Rule dephasing
rate of Ref.~\cite{Bertet05c}, and is similar to the result of Ref.~\cite{Blais04}. These previous results have been obtained considering only the two-point correlator of the fluctuating observable $(a+a^{\dagger})^2$, i.e.~assuming an Gaussian environment. The crossover point $\kappa_m$ from $1/\kappa$ to $\kappa$ in
Eq.~(\ref{dephasingrate}) is, at the Delft parameters \cite{Bertet05b}, comparable to
$\Omega$, i.e., $\kappa$ would dominate over $1/\kappa$ only in a
regime where the Born approximation fails.  Nevertheless, since the golden rule limit $\lim_{\kappa\to\infty}1/\tau_{\phi}=0$ is unphysical, such a term was to be expected.

In the WQOC regime, the enhancement of dephasing by weak coupling to the environment
is analogous to the enhancement of spontaneous emission by the narrow cavity lines in the
{\em resonant} Purcell effect, see Refs.~\cite{Purcell46, Haroche89}. 
In the pure dephasing case we have no energy exchange between the qubit and the oscillator. Qubit  decoherence is caused by fluctuations of $(\hat{a}+\hat{a}^{\dagger})^2$. Since we are in the WQOC regime, the stronger coupling between the oscillator and the environment causes equilibrium between the oscillator and the bath on a shorter time scale than the qubit dephasing.  In equilibrium, the main contribution to the fluctuations of $(\hat{a}+\hat{a}^{\dagger})^2$ is the exchange of photons between oscillator and bath. The process is analogous to equilibrium fluctuations in canonical thermodynamics. A virtual photon returning from the environment is at resonance with the oscillator. The absorbtion of this photon, like in the resonant Purcell effect will be enhanced by narrow oscillator lines. Therefore, the entire dephasing process will be enhanced when the coupling to the environment is weak { and this mechanism can be viewed as a phase Purcell effect. We give a more detailed discussion of this effect in the Appendix.}

\section{Strong qubit-oscillator coupling} 

The dephasing rate
(\ref{dephasingrate}) obtained in the small $\kappa$ and WQOC limit
diverges for $\kappa\to0$, i.e., in the absence of an environment. The
solution to this apparent contradiction lies beyond the WQOC, therefore we solve Eq.~(\ref{offdiag}) numerically using again the Gaussian ansatz for $\chi_{\uparrow\downarrow}$.

\begin{figure}[h]
  \includegraphics[width=.45\textwidth]{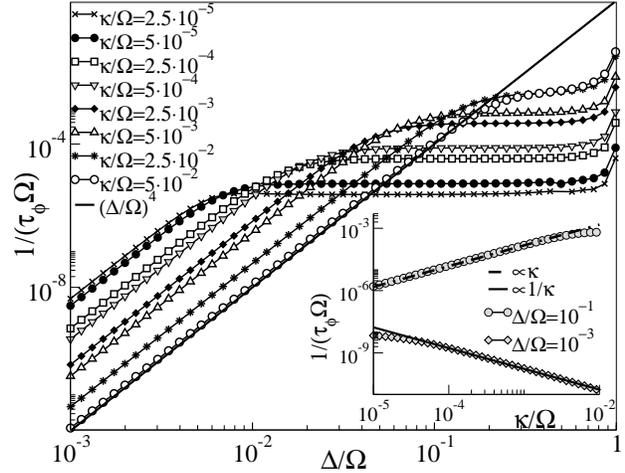}
  \caption{Dephasing rate $1/\tau_\phi$  as function of $\Delta$ for
different values of $\kappa$. Power-law $\Delta^4$ growth at low
$\Delta$ crosses over to $\Delta$-independence at strong
coupling.  Inset: Dephasing rate as a function of $\kappa$ in the
weak coupling regime ($\Delta/\Omega=10^{-3}$) showing
$1/\tau_\phi\propto\Delta^4/\kappa$ and the strong coupling
regime ($\Delta/\Omega=10^{-1}$) with $1/\tau_\phi\propto\kappa$
dependence.  Here $\hbar\Omega/k_BT=2$ similar to
experiments.}\label{gamma}
\end{figure}

Figure~\ref{gamma} shows the dependence of the dephasing rate
on, $\Delta$ for various values of $\kappa$.  The
dimensionless parameter $\hbar\Omega/k_BT$ is $2$, similar to the Delft and Yale setups.
As predicted by eq.~(\ref{dephasingrate}) for $\kappa\ll\kappa_m$ and
small $\Delta$, the dephasing rate is proportional to
$\Delta^4/\kappa$. Further increasing $\Delta$, we observe a saturation of the dephasing rate which marks the onset of the strong coupling regime. This regime is analogous to the strong coupling in linear cavity QED. Here $1/\tau_\phi$ is proportional to $\kappa$.

\begin{figure}[h]
  \includegraphics[width=.45\textwidth]{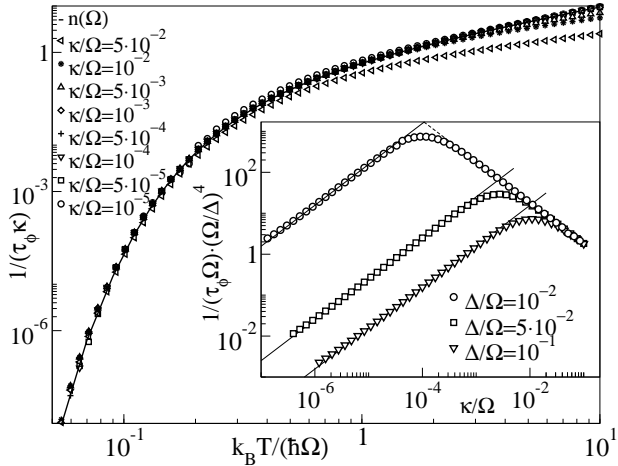}
  \caption{Scaling plot of the dephasing rate $1/\tau_\phi$ as
function of temperature. For $\Delta/\Omega=0.3$, i.e. in the strong
coupling regime (see Fig.~\ref{gamma}), for a wide range of
$\kappa$'s we show that $1/(\tau_\phi\kappa)$ is proportional to the
Bose function $n(\Omega)$. Inset: dephasing rate $1/\tau_\phi$  as
function of $\kappa$ for different values of $\Delta$ and
$\hbar\Omega/k_BT=2$. Continuous lines correspond to $\kappa
n(\Omega)$, dashed lines correspond to
$n(\Omega)(n(\Omega)+1)\Delta^4/(\Omega^2\kappa)$.}\label{knt}
\end{figure}

At strong qubit-oscillator coupling the oscillator couples to
the qubit stronger than it couples to the heat bath, such that one
cannot use the effective bath concept of WQOC. As the
qubit-oscillator system becomes entangled, a fundamentally different
dephasing mechanism emerges. The eigenstates of the Hamiltonian
(\ref{eq:Hamiltonian}) at $\kappa=0$ are the dressed states $\{|\sigma,m_\sigma\rangle\}$ where $|m_\sigma\rangle$ are the number states of the oscillator with frequency $\Omega_\sigma$. Opposed to WQOC, these dressed states are built in the strong coupling regime on a shorter time scale than the re-thermalization of the oscillator. In the evolution from thermal state of oscillator with frequency $\Omega$ to an equilibrium between the new oscillator with $\Omega_\sigma$ and the bath, the state in the narrower potential tends to absorb and the one in the wider potential to emit photons to the bath in an incoherent manner, causing fluctuations of $(\hat{a}+\hat{a}^\dagger)^2$ and thus qubit decoherence.  Thus we expect $1/\tau_\phi\propto \kappa n(\Omega)$. This simple picture is confirmed by
numerical results in Fig.~\ref{knt}, for a  wide range of values of
$\kappa$.
The inset of Fig.~\ref{knt} shows the crossover from strong coupling
rate $\kappa$ to WQOC rate $1/\kappa$. This indicates that, for fixed
$\Delta$, as $\kappa$ decreases, $\Delta$ stops being "small" and
the WQOC limit breaks down. Thus, approaching $\kappa=0$ for any given $\Delta$ we
eventually leave the domain of validity for eq.
(\ref{dephasingrate}) avoiding the divergence at $\kappa\to0$. As
expected, dephasing will vanish as we go to a finite quantum system
(qubit $\otimes$ single oscillator) at $\kappa=0$. We observe that
the criterion of "small" $\Delta$ in WQOC is valid only relative to
$\kappa$.  Using $1/\tau_{\phi}=\kappa n\!\left(\Omega\right)$
in the strong coupling regime and the $1/\kappa$ term of
$1/\tau_{\phi}$  in eq.~(\ref{dephasingrate}) in the weak coupling
regime, we determine the position of the crossover $\Delta_c$
between the two regimes
\begin{eqnarray}
\frac{\Delta_c}{\Omega}&=&\sqrt{\frac{\kappa}{\Omega}}\frac{1}
{(1+n(\Omega))^{1/4}} \,\, .
\end{eqnarray}
The position of the cross-over is controlled by the ratio of the
coupling strengths between the three subsystems i.e.~$\Delta^2/\Omega$ and $\kappa$. Note that, with the
in-situ tuning of the qubit-SQUID coupling, available in the Delft
experiment, the position of the cross-over could be tested
experimentally. Using the parameters from Ref.~\cite{Wallraff04,
Bertet05b} one finds  $(\Delta/\Delta_c)_{\rm
Yale}\approx1.4$ and $(\Delta/\Delta_c)_{\rm
Delft}\approx1.3$ i.e., the strong coupling regime finds application in both setups .

If the oscillator is weakly driven off-resonance, as is the case in the dispersive
measurement, the qualitative behavior remains the same as in
Fig.~\ref{gamma}, as shown in Ref. \cite{Serban06}. In
general a tunneling $\hat{\sigma}_x$ term may occur in
Eq.~(\ref{eq:Hamiltonian}) and lead to energy relaxation as well as further
reducing the matrix elements containing the dephasing rate. We
expect that, as long as the energy splitting $E$ of the qubit is
off-resonance with the oscillator, which in our case means
$|E-2\Omega|\gg \kappa$, the effect of the relaxation is rather weak
and dephasing still dominates. On resonance, we expect a similar
Purcell to strong coupling crossover as for the dephasing channel.

Our results  have applications in other systems with similar dispersive qubit-oscillator coupling, e.g., the Yale setup \cite{Wallraff04}, in the off-resonant dispersive regime.
There, the system is described by a similar (Eq.~(12) in Ref.~\cite{Blais04})
quadratic coupling $\hat{a}^{\dagger}\hat{a}$ between qubit and
cavity and a pure dephasing Hamiltonian. In particular, a strong
dispersive regime of this system has been utilized to resolve number
states of the electromagnetic field in Ref.~\cite{Schuster06}. The
terms $\hat{a}^2$ and $\hat{a}^{\dagger2}$ in Eq.~(1) do not play a
central role for our physical predictions, as confirmed by the numerical simulations. We expect our results, with minor adaptations, to
be applicable to various cavity systems, e.g. quantum dot or atom-based
quantum optical schemes \cite{Raimond01,Balodato05}. The dispersive coupling of Hamiltonian (\ref{eq:Hamiltonian}) could have implications for the generation of squeezed states, quantum memory in the frame of quantum information processing, measurement and post-selection of the number states of the cavity. 
\section{Conclusion}
{ We have presented a concise theory of the dephasing
of a qubit coupled to a dispersive detector spanning both strong and weak coupling. The phase-space method applied is based on treating the oscillator on the same level of accuracy as the qubit. We have discussed the dominating decoherence mechanism at  weak qubit-oscillator coupling, where the linewidth of the damped oscillator plays the main role, analogous to the Purcell effect.  At strong qubit-oscillator coupling we have identified a qualitatively different behavior of the qubit dephasing and discussed it in terms of the onset of the qubit-oscillator entanglement. We have provided a criterion delimitating the parameter range at which these processes dominate the qubit dephasing.}

\acknowledgements
We are especially grateful to F.~Marquardt, I.~Chiorescu and A.~Blais for very
helpful discussions.  We thank J.v.~Delft, A.G.~Fowler and
C.~Couteau for many useful suggestions. This work is supported by
NSERC and the DFG through SFB 631 and by EU through EuroSQIP and
RESQ projects. I.S.~acknowledges support through the Elitenetzwerk
Bayern.
\section{Appendix}
Assuming the WQOC limit we use Fermi's golden rule in an otherwise exact manner to prove the analogy between the weak qubit-oscillartor coupling regime and the Purcell effect. One can map the damped oscillator by an exact normal mode transformation
\cite{Ambegaokar05} onto an \emph{effective} heat bath of
\emph{decoupled} oscillators denoted by
$\hat{c}_j,\hat{c}_j^\dagger$ and with a spectral density
\begin{equation} J_{\rm
eff}(\omega)=\frac{2\kappa\omega}{(\omega^2-\Omega^2)^2+\kappa^2\omega^2}.
\label{eq:jeff}
\end{equation}
$J_{\rm eff}$ corresponds to the effective density of electromagnetic modes in the cavity introduced
in regular linear cavity QED for describing the Purcell effect.
The WQOC decoherence rate is proportional to  the two-point correlation function of the
environmental operator coupling to the qubit \cite{Slichter96,Nato06II}, in our case
\begin{equation} S_2(\omega)=\left\langle
\hat{X}^2(t)\hat{X}^2(0)\right\rangle_\omega-\left(\langle
\hat{X}^2\rangle\right)^2,\label{corr}
\end{equation} where  $\hat{X}$ is the sum of the \emph{effective}
bath coordinates $\hat{X}=\sum_j
\sqrt{\hbar/(2m_j\omega_j)}(\hat{c}_j+\hat{c}_j^\dagger)$.  For
the pure dephasing situation described by the Hamiltonian
(\ref{eq:Hamiltonian}) we only need to study $1/\tau_{\phi}\propto S_2(\omega\to0^+)$ because the qubit energy conservation implies energy conservation within its effective environment.  The last term
of Eq.~(\ref{corr}) removes the noise bias.  This is important since
dephasing is caused only by processes that leave a trace in the bath
\cite{Leggett02}, i.e.~the exchanged boson spends a finite time in the environment.
Terms of the structure 
$\left\langle\hat{c}^\dagger_i(t)\hat{c}_j^\dagger(t)\hat{c}_k \hat{c}_l\right\rangle$ ,$\left\langle\hat{c}_i(t)\hat{c}_j(t)\hat{c}_k^\dagger \hat{c}_l^\dagger\right\rangle$ contribute to $S_2(0)$ only when $\omega_i=\omega_j=0$, which are modes with density $J_{\rm eff}\simeq2\kappa\omega/\Omega^4$ each, leading to terms are of order $\kappa^2$.
Up to linear order in $\kappa$, the only terms in $S_2(\omega\to0^+)$ that fulfill the energy conservation and leave a trace in the bath are
of the structure $\left\langle
\hat{c}^\dagger_l(t)\hat{c}_j(t)\hat{c}_j^{\dagger} \hat{c}_l
\right\rangle$, including the permutations among the operators taken at
time $t$ and those taken at time 0. The terms contributing to $S_2(\omega\to0^+)$
satisfy the condition $|\omega_l-\omega_j|\to 0^+$. Physically this corresponds to infinitesimal energy fluctuations which leave a trace in the bath. Or, in other words, the photon absorbed at t=0, $\hat{c}_l$, should spend finite time in the bath and be emitted back only at the later time t, but at the same time the energy change in the environment e.g. caused by $\hat{c}_j^{\dagger}\hat{c}_l$ should remain undetectable within the energy-time uncertainty at every time, therefore in the Golden Rule (long time) limit $\omega_l\approx\omega_j$. Taking the continuum limit
we thus have 
\begin{equation} 1/\tau_{\phi}\propto\int_0^{\infty}\! \!d\omega\: J_{\rm
eff}(\omega)(1+n(\omega))J_{\rm
eff}(\omega)n(\omega)\label{continuum}.
\end{equation}
The integral in eq.~(\ref{continuum}) can be  rewritten as the convolution
$K(\omega')=\int d\omega J_{\rm
eff}(\omega)n(\omega)J_{\rm eff}(\omega'-\omega)n(\omega'-\omega)$
for $\omega'\to0$. Using eq.~(\ref{eq:jeff}), $K(\omega')$ becomes a
function with resonances at $\omega'=0$ and $\omega'=2\Omega$. The
the height of these resonances and consequently $1/\tau_{\phi} \propto S_2(0)$ increases with decreasing $\kappa$, thus matching the behavior of the dephasing rate (\ref{dephasingrate}). 
At the same time, the tail of the peak at $2\Omega$ enhances $S_2(0)$ when $\kappa$ increases. This corresponds to the $\kappa$ term in eq. (\ref{dephasingrate}).
Analogous to $1/\tau_\phi$ in eq.~\ref{dephasingrate}, $S_2(\omega\to0^+)$ vanishes for $T\to0$.

\end{document}